\documentclass[aps,pra, twocolumn]{revtex4-1}
\usepackage{amsthm}

\usepackage[utf8]{inputenc}
\usepackage[sc,osf]{mathpazo}

\usepackage{graphicx,subfigure}
\usepackage{bm}
\usepackage{makeidx} 
\usepackage{amsmath}
\usepackage{amssymb}
\usepackage{color}
\usepackage{amsthm}
\usepackage{hyperref}
\usepackage{cleveref}
\usepackage{diagbox}
\hypersetup{colorlinks, citecolor=magenta, filecolor=blue, linkcolor=blue, urlcolor=green}
\usepackage[normalem]{ulem}
\usepackage{float}
\usepackage{multirow}

\newcolumntype{P}[1]{>{\centering\arraybackslash}p{#1}}
\begin{document}

\title{Decoherence-free mechanism to protect  long-range entanglement  against  decoherence}

\author{Leela Ganesh Chandra Lakkaraju, Srijon Ghosh, Aditi Sen (De)}

\affiliation{Harish-Chandra Research Institute, HBNI, Chhatnag Road, Jhunsi, Allahabad 211 019, India}

\begin{abstract}

Quantum spin models with variable-range interactions can exhibit certain quantum characteristics that a short-ranged model cannot possess. By considering the quantum XYZ model whose interaction strength between different sites varies either exponentially or polynomially, we report the creation of long-range entanglement in dynamics both in the absence and presence of system-bath interactions. Specifically, during closed dynamics, we determine a parameter regime from which the system should start its evolution so that the resulting state after quench can produce a high time-averaged entanglement having low fluctuations. Both in the exponential and  power-law decays, it occurs when the magnetic field is weak and the interactions in the z-direction are nonvanishing. When part of the system interacts with the bath repeatedly or is attached to a collection of harmonic oscillators along with dephasing noise in the z-direction, we observe that  long-range entanglement of the subparts which are not attached with the environment remains constant with time in the beginning of the evolution, known as freezing of entanglement, thereby demonstrating a method to protect long-range entanglement. We find that the frozen  entanglement content in any length and the time up to which freezing occurs called the freezing terminal to follow a complementary relation for all ranges of interactions. However, we find that for a fixed range of entanglement,  there exists a critical value of interaction length which leads to the maximum freezing terminal.  

\end{abstract}

\maketitle

\section{Introduction}

In the path of advancement in second-generation quantum technologies, quantum entanglement is shown to be indispensable \cite{ent1} in tasks ranging from quantum communication  with and without security \cite{dc, tele1, crypto1, crypto2} to  measurement-based quantum computation \cite{oneway1, oneway2, oneway3, oneway4, oneway5}. Over the last two decades,  by using the concepts from quantum information science \cite{ultracold-review, fazio-rev}, several interesting questions in many-body physics are addressed which include scaling of entanglement at quantum critical points \cite{qptbook2, qpt, qptfazio},  developing methods for finding the ground state of a Hamiltonian \cite{mpsrev},  detecting  non-equilibrium phenomena like dynamical quantum phase transition \cite{Heylrev, stavPRB, stavPRR}, to name a few. 
On the other hand, remarkable progress in the field of atomic, molecular and optical systems, including trapped ions \cite{ionrev},  photonic systems \cite{revphotons}, cold atoms in optical lattices \cite{revatoms, coldrev} and  superconducting circuits \cite{superc1, superc2}, give rise to the possibility of realizing and manipulating quantum many-body systems in a controlled manner, and hence can  generate entangled state in physically realizable systems with high fidelity in laboratories. Moreover, entanglement in dynamical states of quantum spin models which are created by suitably tunning the interaction strengths between the subsystems and other  relevant parameters turn out to be  useful resouces in measurement-based quantum computation \cite{briegel2009measurement}, quantum state transfer \cite{Sougato}, quantum thermal machines like quantum refrigerators \cite{Qrefri}, quantum batteries  \cite{battrev}. 

Recent experiments in ion traps \cite{Neyenhuise1700672, Islam583, Jurcevic2014} and other physical systems also demand to probe the  physical properties of interacting quantum spin systems with variable range interactions \cite{RevModPhyspower, csb, lahaye2009physics, CSB1, CSB2}. 
At the same time, long-range quantum spin systems can exhibit a rich phase diagram which cannot be seen in  short-range models. Specifically,  it was found that Heisenberg long-range models with power-law decay possess a continuous symmetry breaking phase along with ferromagnetic, XY and antiferromagnetic ones. Therefore, it is interesting to find out whether the model can provide an interesting platform for generating entanglement, thereby making this model lucrative for quantum technologies.  Till date, all the investigations of quantum information theoretic quantities  in these models  have been carried out to explore the static properties  \cite{PhysRevA.98.023607, PhysRevB.101.094410, lakkaraju2020distribution}.

In this manuscript, we go beyond it and scrutinize  quantum correlations of the evolved state when the initial state is the canonical equilibrium state of the anisotropic quantum XYZ model in presence of a uniform magnetic field in the $z$-direction with variable range interactions following a  power-law and exponential-law decays. For evolution, the quenching is performed by switching-off the magnetic field. We identify the parameter-regimes which are admissible to tune, so that high time-averaged entanglement, quantified by logarithmic negativity \cite{neg4, neg5} is produced with low fluctuations which we measure via the standard deviation of entanglement. In particular,  we find that in presence of a weak magnetic field, the XYZ model  leads to the high amount of nearest neighbor as well long-range entanglement  production on average compared to that of the XY model  although the regimes giving high averaged entanglement  also pay a cost of high fluctuations.
Moreover, we report that the advantage of the long-range model is eradicated with the increase in strength of interactions for the power-law decay. 

During implementations of quantum protocols, the most common hindrances occur due to the errors in manipulation  of the system or due to the system-environment interactions,  inescapable in all practical purposes. On one hand, to protect the system from errors, error-correcting codes \cite{shorecode, ecorrcode, nielsenchuang} have been developed, while  several mechanisms have been proposed which can assure the slow rate of decay of quantum properties in open systems. The prominent ones in the later direction include decoherence-free subspace \cite{dfree1, dfree2, dfree3, decohmech}, dynamical decoupling \cite{dyndec1, dyndec2, dyndec3}.  Recently, it was also shown that for a suitable choice of systems and for a certain kinds of noise models,  quantum correlations in the form of entanglement as well as quantum discord \cite{discorev1, discorev2}  can remain constant  for a certain period of time at the beginning of the evolution -- a counter-intuitive phenomenon  is known as freezing of quantum correlations \cite{discofreeze1, discofreeze2, discofreeze3, entfreeze, titfreeze}.  All the above studies are either restricted to the specific spin system having short-range interactions or a specific form of a state as the initial state of the system. In this respect, it is also important to stress here that the entanglement flow under these decoherence models can be understood from the Lieb-Robinson (LR) bound \cite{kliesch2014lieb} which are well understood in the short-range model (for variable-range LR bound, see \cite{nachtergaele2011liebrobinson, hastings2006spectral} for exponential decay and \cite{Sweke_2019, PhysRevX.10.031009} for power-law decay). 

We propose here a decoherence-free set-up -- a system consists of a few spin-$1/2$ particles, described by the XYZ Hamiltonian  having variable-range interactions in presence of a magnetic and a part of the system is affected by noise. The environment  is modeled either by the collection of thermal states interacting individually with the subparts of the systems for a certain small period of time repeatedly \cite{dhahri2008lindblad, PhysRevLett.102.207207, PhysRevA.91.040303, attal2006repeated, PhysRevA.34.1642, barchielli1991measurements} or by  local bosonic baths and the local dephasing noise \cite{noise1, PhysRevLettbosonic} in the \(z\) as well as in the \(x\) directions.  We observe that in all these situations, both short-range and long-range entanglement  shared between subparts of the entire system can be frozen during a certain period of time, referred to as a freezing terminal. It is maximum  when the modulus of \(zz\)-interactions is weak and the magnetic field is strong or vice versa. We report that to obtain the maximum freezing terminal, there exists a critical range of interactions above which it starts decreasing. 
 Moreover, we report that freezing terminal and the amount of frozen entanglement follow a complementary relation and long-range frozen entanglement is enhanced in presence of variable-range interactions.

This paper is represented as follows. In  Sec. \ref{sec:model}, we describe the static properties of the variable-range XYZ model including phases, the way we quench the system to study the evolved state and time-averaged bipartite entanglement and standard deviation. The time-averaged entanglement and its fluctuations under unitary evolution  are presented in Sec. \ref{sec:close}. In Sec. \ref{sec:open}, by considering the interaction of system with surroundings via repetitive interaction and bosonic bath, we discuss the freezing phenomena of long-range entanglement over time, along with complementarity relations. 
Finally, we conclude in Sec. \ref{sec:conclusion}.

\section{Model and Methodology}
\label{sec:model}

First we briefly discuss the static properties of the quantum spin model  which include the  phase diagram and the fall-off properties of the interaction strength between the subsystems. We then describe the quenching method that we use here to investigate the evolution of the said model.  We finally introduce  quantities based on bipartite entanglement measure, namely logarithmic negativity \cite{neg4, neg5}  for studying the dynamics of entanglement.

\subsection*{Spin models: Statics vs. Dynamics}

The Hamiltonian describing the anisotropic quantum $XYZ$ Heisenberg spin model  consisting of $N$ spin-$\frac{1}{2}$ particles with variable-range interactions having open boundary
 conditions reads as
\begin{eqnarray}
H &=& \sum_{\substack{i < j}}^{N} J_{ij} \Big[ \frac{1+\gamma}{4}\sigma_i^x\sigma_j^x + \frac{1-\gamma}{4}\sigma_i^y\sigma_j^y \Big]  + \frac{\Delta'_{ij}}{4}\sigma_i^z\sigma_j^z \nonumber \\ &+& \sum_{i=1}^N \frac{h}{2} \sigma_i^z,
\label{eq:ham1} 
\end{eqnarray}
where $\sigma^k, \, k=x, y, z$ is the Pauli spin matrices, $J_{ij}$  and $\Delta'_{ij}$ are respectively  the coupling  constants along the  $x-y$ plane and in the $z$-direction,  $\gamma$ is the anisotropy parameter,  and $h$  the strength of the  magnetic field in the transverse direction. We consider all possible interactions between the spins, so that $i$ runs from $1$ to $N-1$ and $j$ runs from $i+1$ to $N$.  As we are interested to probe the physics of variable-range scenario, we consider two qualitatively different fall-off of the interaction strength
 between spin $i$ and  $j$, namely, exponential decay with $J_{ij} \sim \alpha_e^{-(|i-j|-1)}$ and power-law decay where $J_{ij} \sim |i-j|^{-\alpha_p}$ with  fall-off rates of the interaction strengths for the exponential and power-law decays  respectively being $\alpha_{e (p)}$. Hence, coupling in the $x$ - $y$ plane can be rewritten as
$\frac{J_{ij}}{J} = \alpha_e^{-(|i-j|-1)} \text{ or } |i-j|^{-\alpha_p}$,
where  $J$ is a constant and of a ferromagnetic-type, i.e., $J<0$. Similarly,
$\frac{\Delta_{ij}}{\Delta} = \alpha_e^{-(|i-j|-1)} \text{ or } |i-j|^{-\alpha_p}$,
where \(\Delta\) can be both positive as well as negative, i.e., both ferromagnetic and antiferromagnetic in nature. The reason behind considering open boundary is that in the study of  open quantum dynamics,  we want to minimize the effect of environment on the properties of the system. To make the parameters dimensionless, we fix  $h/|J| = \lambda$, and  \(\Delta'/|J| = \Delta \).

By varying \(\alpha_p\), the rich phase diagram of the above model has been studied when \(\gamma =0\) \cite{csb}. By using bosonization and density matrix renormalization group techniques \cite{qptbook2}, it was found that the model posses  phases like continuous symmetry breaking (CSB) \cite{CSB2, CSB1}, XY, ferromagnetic and antiferromagnetic in the plane of \(\Delta\) and  \(\frac{1}{\alpha}\). Specifically, when \(\Delta<0\), by varying \(\alpha_p\), one obtains a transition from CSB to XY while for antiferromagnetic coupling in the z-direction, CSB to antiferromagnetic transition occurs. Our aim in this paper is to initially prepare the system in a specific phase, and by quenching the magnetic field, the generation of entanglement in the evolution is studied, thereby identifying the possible parameters which are beneficial for entanglement-production. Moreover, we vary the coordination number of the model, defined as \(|i-j|\leq \mathcal{Z}\), so that the effects of \(\mathcal{Z}\) on entanglement creation can be understood. 
 

\emph{Dynamics of isolated system.} Let us now move to the evolution of the system. The initial state is taken to be the  canonical equilibrium state of the Hamiltonian, denoted by $\rho^{\beta}(t=0)$. For an inverse temperature $\beta = \frac{1}{K_{B}T}$, with $K_B$ being the Boltzman constant and $T$ being the absolute temperature, the initial $N$-party state reads as
\begin{equation}
\label{eq:thermal}
\rho^{\beta} (t=0) = \frac{e^{-\beta H(\lambda \ne 0,\gamma,J,\Delta, \alpha_{p (e)})}} {\text{tr} (~e^{-\beta H})}.
\end{equation}
We then quench the system by switching off the magnetic field, i.e.,  $h = 0$, so that the time-evolved state obtained via unitary dynamics is given by \(\rho (t) = U \rho^\beta (t =0) U^{\dagger} \) with \( U = e^{-i t H(\lambda = 0,\gamma,J,\Delta, \alpha_{p (e)})}\) and \(\lambda = a, \; t \leq 0, \); \(\lambda = 0, \; t > 0\). To make parameters independent of dimension, we call \(\beta/|J|\) and \(t/|J|\) as \(\beta\) and \(t\) respective;y. 
%



Since the model cannot be solved analytically, we resort to numerical diagonalization technique. Moreover, to investigate entanglement, we require all the correlations and magnetizations of any two-party density matrices of the evolved state. Since we will also later deal with open system dynamics of the model, we 
 stick to a moderate size of the system, i.e.,  $N=8$. In order to mitigate the boundary errors, we start evaluating entanglement between the two neighboring spins at the centre of the chain, i.e., $(4,5)$-pair whose density matrix is denoted by \(\rho_{45}\) and then to study the behavior of entanglement length, we calculate entanglement in the \((4, 5+r)\)-pair, with \(r =1,2,3\). 

\subsection*{Time-averaged entanglement and its standard deviation }


A notable bipartite entanglement measure for two spin-$1/2$ particles is logarithmic negativity (LN), \(\mathcal{L}\), \cite{neg4, neg5}, which is  a function of negativity, $ {\cal N} =\sum_i |e_i|$, with \(e_i\) being the negative eigenvalues of the partial transposed state with respect to any of the party. LN  for a bipartite state, \(\rho\) is then given by \(\mathcal{L} (\rho) =  \log_2 [2 {\cal N} + 1].$ To find out the behavior of bipartite entanglement between a pair, \((i,j)\), we construct the density matrix of dimension $2^N\times2^N$ obtained from the Hamiltonian of \(N\) spins,  trace out all the parties except \(i\) and \(j\) and investigate the properties of   \(\mathcal{L} (\rho_{(i,j)}(t))\) as a function of other system-parameters. For notational simplicity, round parenthesis and comma for marking the pair will be omitted when both are numbers. 

Typically, we observe that  bipartite entanglement of any pair starting from a nonvanishing value collapses and revives with time. For some \(h\) value, we sometimes do not get any nonvanishing entanglement at large time. To identify the parameters of the Hamiltonian which can lead to high entanglement on average during the dynamics, 
we study two moments of entanglement, mean and standard deviation of entanglement in dynamics, which quantify the entanglement that can be produced on average as well as the fluctuations of entanglement with time. 
The average  entanglement created during the evolution, namely the time-averaged entanglement between any pair, can be defined as 
\begin{equation}
\label{eq:avgent}
\mathcal{L}_{avg} = \frac{1}{n} \sum_k \mathcal{L} (\rho(t_k)),  
\end{equation}
where \(t_k\) is the each instance of time in which entanglement is calculated from the evolved state, \(\rho\), and  \(n\) is the total duration of the dynamics, i.e.,   $n = \frac{t_f-t_{in}}{t_s}$, with  $t_f$, $t_{in}$ and  $t_s$  respectively being the final, initial, and the incremental time.
Typically, we perform averaging from the initial time till \(t_f  = 200\) and \(t_s =0.01\), so that \(n =20000\). 
In order to observe the entanglement-fluctuations over time, we use standard deviation, given by
\begin{equation}
\mathcal{L}_\sigma = \sqrt{{\frac{\sum_k (\mathcal{L} (t_k) - \mathcal{L}_{avg})}{n}}^2}.
\end{equation}
In case of nearest neighbor spin model, entanglement is present only in the nearest neighbor (NN)  pair and next-nearest neighbor (NNN) pair while in the variable-range interacting model in Eq. (\ref{eq:ham1}), we will show that long-range entanglement is present, not only its initial thermal state, but also in the time-evolved state.

 \begin{figure}[h]
 \includegraphics[width = 8.6 cm]{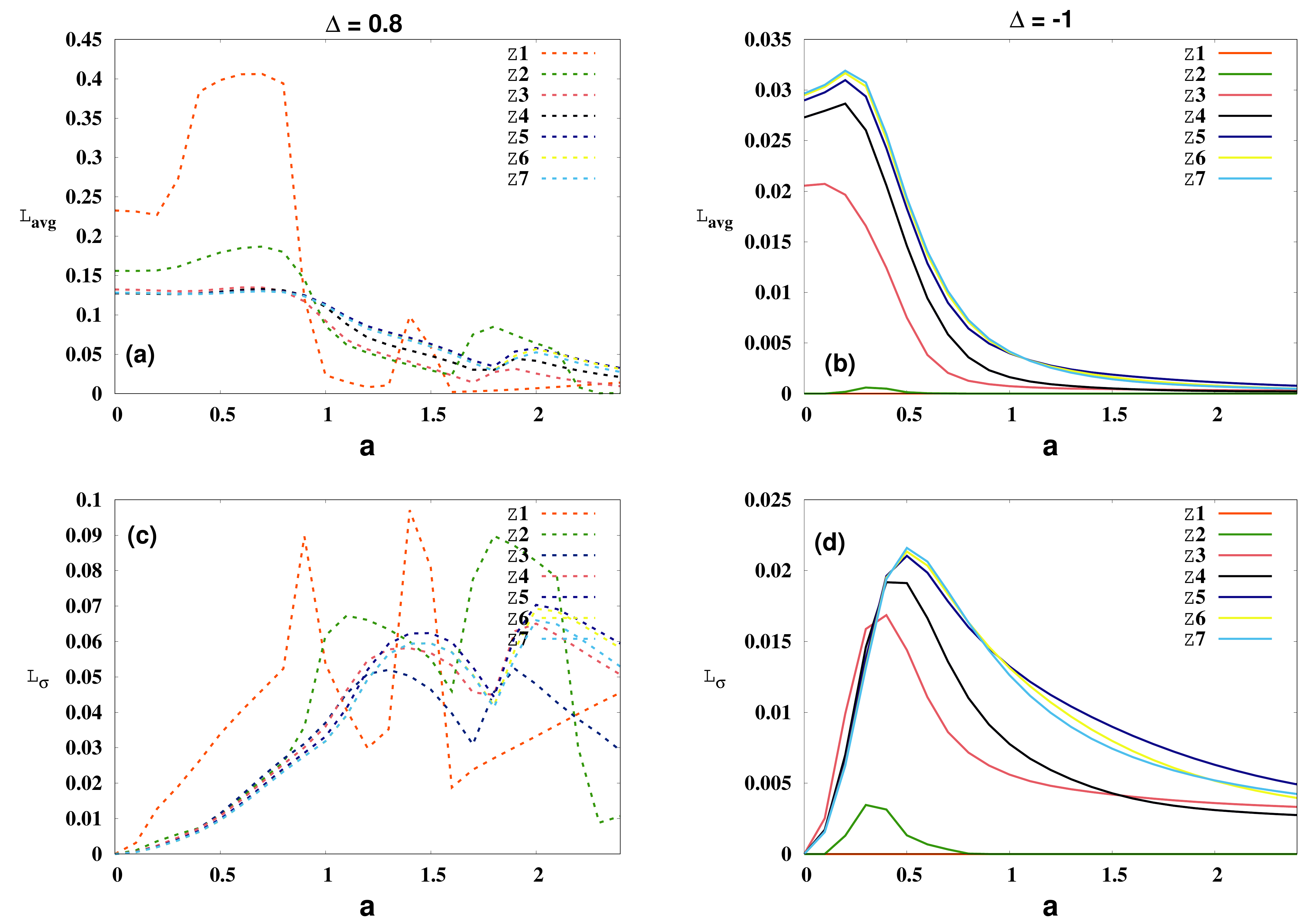}
 \caption{(Top  panel) Nearest neighbor time-averaged entanglement, $\mathcal{L}_{avg} (\rho_{45})$ (ordinate)  against \(a\) (abscissa).  (Bottom panel) The dispersion, $\mathcal{L}_{\sigma}$ is  against \(a\). The initial state is chosen to be the canonical equilibrium state of the Hamiltonian (with open boundary condition), given in Eq. (\ref{eq:ham1}) with power-law decay,  \(\alpha_e =2\). Here \(N =8\), \(\gamma =0.8\) and \(\beta/|J| =200\).   The number next to \(\mathcal{Z}\) represents the range of interactions, i.e., "\(2\)" corresponds to the NNN interacting Hamiltonian while "\(7\)" is the long-range Hamiltonian. Left panels are for \(\mathcal{Z} =2\) while the right ones are with \(\mathcal{Z} =7\). The ordinates are in ebits and abscissae are dimensionless.    }
 \label{fig:45vsaexp}
 \end{figure}  

 \begin{figure}[h]
 \includegraphics[width = 8.6 cm]{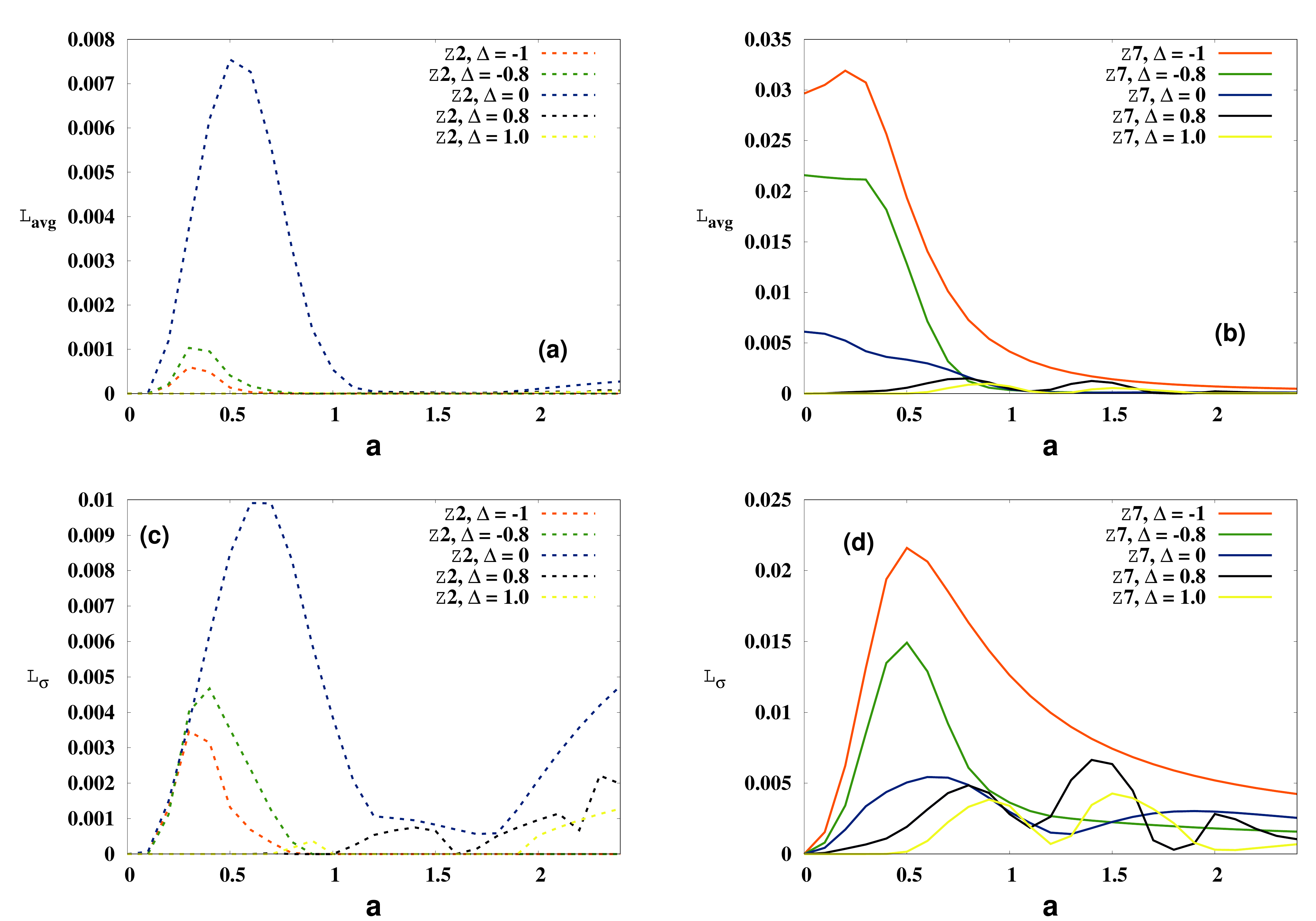}
 \caption{(Left panel) Time-averaged entanglement, $\mathcal{L}_{avg} (\rho_{47})$ (on top) and  $\mathcal{L}_{\sigma} (\rho_{47})$   (bottom) against \(a\) with   \(\alpha_e =2\). All the configurations are same as in Fig. \ref{fig:45vsaexp}. }
 \label{fig:47vsaexp}
 \end{figure}

\section{Closed dynamics of average  long-range entanglement and their fluctuations} 
\label{sec:close}
In this part, we will study the time-averaged entanglement between different pair produced in evolution.  Both \(\mathcal{L}_{avg}\) and \(\mathcal{L}_\sigma\) are functions of \(\lambda\), \(\Delta\),  \(\alpha_{e (p)}\), \(\beta\) and \(\mathcal{Z}\). In the entire analysis, we keep temperature to be large enough so that the thermal fluctuations in the system is minimized. Unless mentioned otherwise, we fix \(\beta =200\). Below, we first discuss the situation when the interaction strength follows the exponential decay and then we move to the case of power-law decay.

\subsection{Entanglement dynamics with exponential-decay}

Let us first deal with the scenario when  each spin interacts with its neighboring spins having exponentially decaying interaction strengths, it is fixed to  $\alpha_e = 2$ so that the interaction drops as 
\begin{equation}
J_{ij} \sim 2^{-(|i-j|-1)}
\end{equation}
In this setting, we investigate the behavior of entanglement by varying the initial strength of the magnetic field, \(a\). Our aim is to find out the effects of \(\Delta\) and \(\mathcal{Z}\) on the creation of entanglement in this situation.

\begin{enumerate}
\item  \emph{Effect of $\Delta$ on NN entanglement.} According to the trends of entanglement, there can be three regions that may play an important role here, namely \(\Delta =0\) representing the XY model, \(\Delta>0\), i.e., the antiferromagnetic \(zz\)-interaction, and \(\Delta <0\). Moreover, the behavior of NN entanglement of  $(4,5)$-pair is also drastically different when the initial magnetic field is small i.e., when \(a <1\) and when \(a >1\).  
We find that for low values  \(a\), the dynamical state typically posses a high amount of entanglement on average in time   with \(\Delta \neq 0\)  compared to the  XY model as depicted in Fig. \ref{fig:45vsaexp}. For high antiferromagnetic interaction in the \(z\) direction, time-averaged entanglement between nearest neighbor pairs reaches their maximum value when \(a \approx 1\)  and increases with the increase of \(\Delta\) while it becomes vanishingly small in presence of high amount of magnetic field, thereby showing some-kind of advantage for moderate to low local magnetic field. On the other hand, patterns of  time-averaged NN entanglement  in presence of ferromagnetic coupling, i.e., \(\Delta <0\),   are qualitatively different. In particular, almost independent of the strength of the external magnetic field,  $\mathcal{L}_{avg}$  is reasonably high with \(\Delta <0\). From this analysis, it is tempting to predict that  to create high amount of NN entanglement on average, we should fix \(\Delta \leq 0.8\) for any range of interactions.  

Before identifying such a parameter space, let us first look at the fluctuations of entanglement in time. Interestingly, we observe that there exists a nice trade off between $\mathcal{L}_{avg}$ and $\mathcal{L}_{\sigma}$.   Specifically, the region in the \(a, \Delta\)-plane which leads to a high amount of time-averaged entanglement, also has high fluctuations with time as shown in Fig.  \ref{stde45vshwithd}. Especially when \(-0.8 \leq \Delta < -1\), $\mathcal{L}_{\sigma}$ is quite high after \( a >1\). Such a trend in $\mathcal{L}_{\sigma}$ can also been seen  moderate to high values of initial magnetic field,  \(a\), irrespective of the variable-range interactions and for all values of \(\Delta\). Therefore, we can conclude that to obtain high NN entanglement in dynamics of this model, we have to choose low \(a\) and high nonvanishing \(zz\)-interactions and any range of interactions. It implies that although \(\Delta\) can give some advantage in production of entanglement in dynamics, \(\mathcal{Z}\) does not play a role in the nearest-neighbor case.

\item  \emph{Long-range entanglement vs. \(\Delta\) and \(a\).} The above analysis leads to an immediate question whether to create long-range entanglement,  \(\mathcal{Z}\) plays any role along with \(\Delta\) and \(a\). Suppose we want to concentrate on entanglement between pairs which is beyond  NN o and NNN, i.e. \(r>1\). Let us consider the entanglement    in the \((4,7)\)-pair of the time-evolved state. As in the previous case, the initial low magnetic field is optimal for achieving high time-averaged entanglement.  In this case, the presence of \(\Delta\) turns out to be related with \(\mathcal{Z}\).    For small values of \(\mathcal{Z}\), XY model produces a maximum amount of entanglement in the \((4,7)\)-pair  while when the interactions are long ranged,   the maximum amount of \(\mathcal{L}_{avg}\) can be found in presence of ferromagnetic moderate \(\Delta\)-coupling. For creating long-range entanglement, non-negative \(\Delta\) turns out to be disadvantageous. Just like in NN case, the fluctuations are high for the system parameters which manages to create  high time-averaged entanglement. 
 
%
 
 
%

 
 
 \begin{figure}[h]
 \includegraphics[width = 8.6 cm]{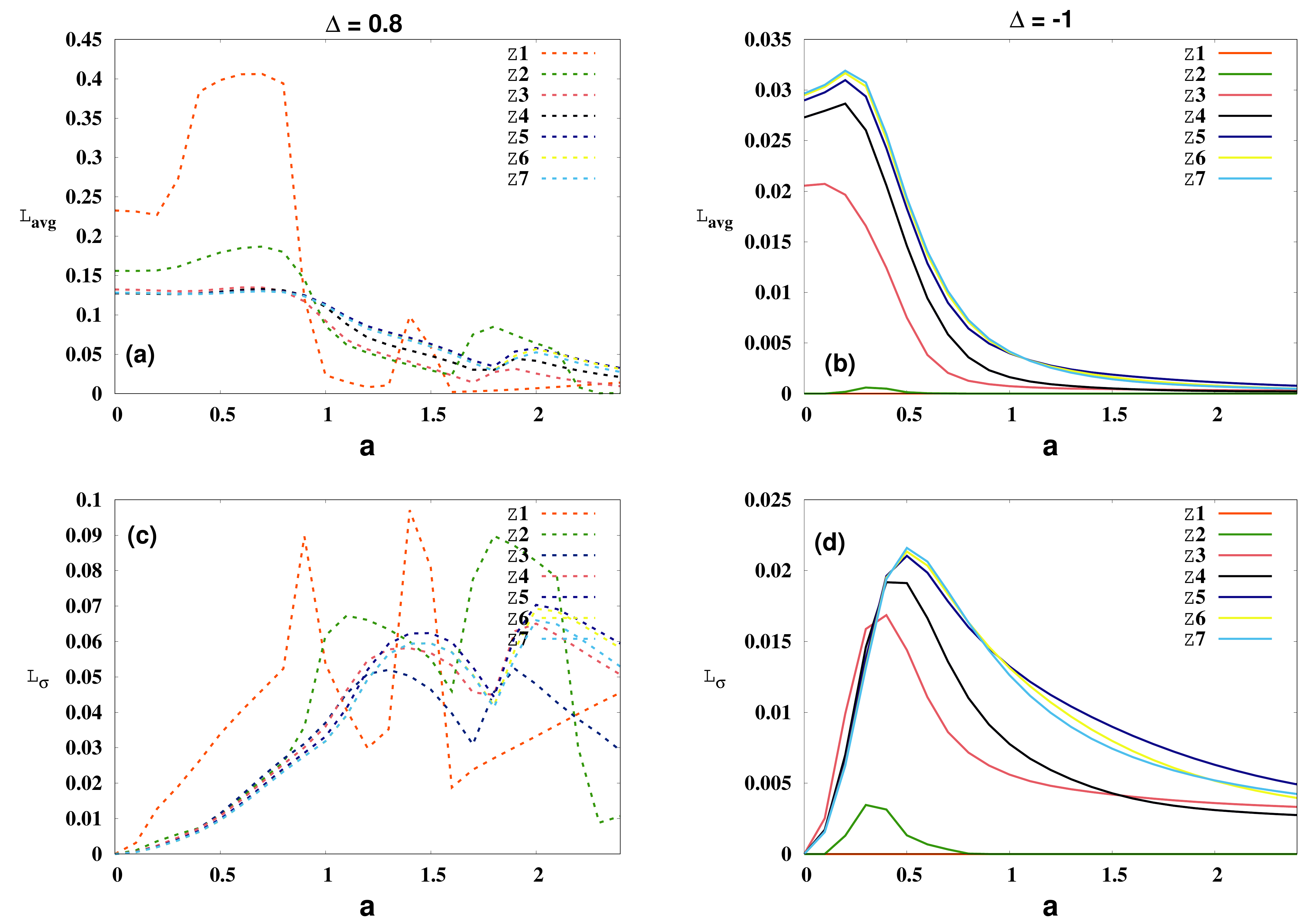}
 \caption{(Left panel) Time-averaged entanglement $\mathcal{L}_{avg} (\rho_{45})$ (top) and $\mathcal{L}_{\sigma} (\rho_{45})$  (bottom) against  \(a\) with different values of \(\mathcal{Z}\).   Here \(\alpha_e =2\) and \(\Delta =0.8\). Right panel is same  as the left one with \(\Delta =-1\). All other configurations are same as in Fig. \ref{fig:45vsaexp}. }
 \label{fig:entwithZexp}
 \end{figure}

 \item \emph{Role of coordination number on entanglement generation. } 
 As observed in the previous situations, the value of $\mathcal{Z}$ has a significant impact on the entanglement values.  
 Starting from the nearest neighbor entanglement, we look for the two-party density matrices, $\rho_{(4,  5+r)}$, ($r = 1, 2, 3$) of the time evolved state. For such investigation, we fix the \(zz\)-interaction strength in a such a way that produces moderate amount of time-averaged entanglement, i.e., we fix \(\Delta = 0.8\). 
After time averaging, we see that $\rho_{45}$ posses a large amount of entanglement when the system is nearest neighbor, i.e., when $\mathcal{Z} = 1$ and \(a\) is small. 
 The picture changes as $\mathcal{Z}$ increases. Specifically, for $\mathcal{Z} \geq 2$, $\mathcal{L}_{avg}$ has higher value compared to $\mathcal{Z} = 1$ in presence of  moderate amount of magnetic field.  
 Clearly,  variable-range interaction is beneficial to create time-averaged entanglement between, say \((4,7)\)-pair. With the increase of \(\mathcal{Z}\), \(\mathcal{L}_{avg}\) increases for any values of initial magnetic field, \(a\).

 
%

%
%
%
\end{enumerate}

  \begin{figure}[h]
 \includegraphics[width = 8.6 cm]{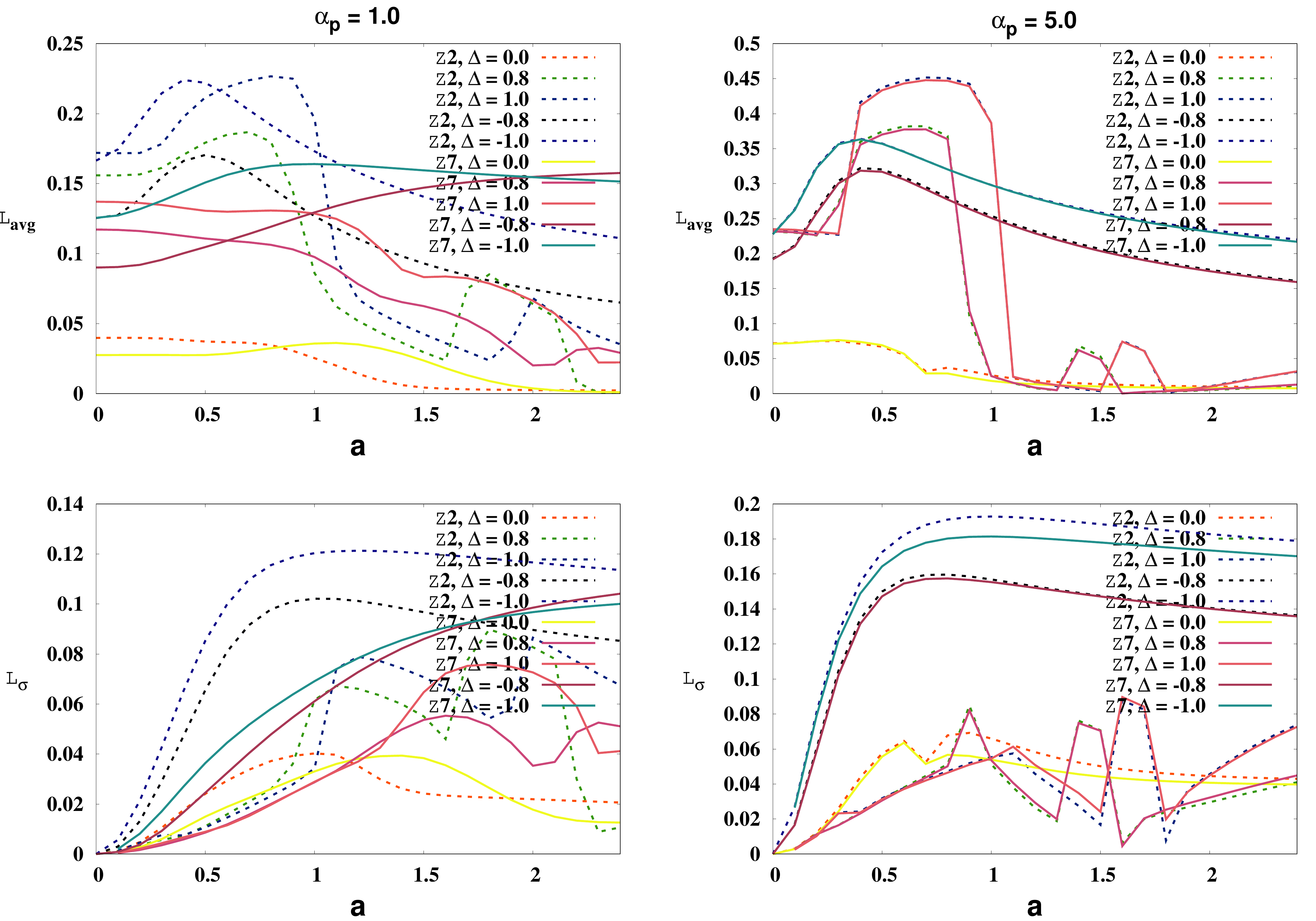}
 \caption{Time-averaged NN entanglement and its standard deviation for power-law decay.  Left panel is for  \(\alpha_p =1\) while the right one is for \(\alpha_p =5\).  We see that the information about  \(\mathcal{Z}\) is erased with the  increase of \(\alpha_p\). All other configurations are same as in Fig. \ref{fig:45vsaexp}}
 \label{fig:45vsa}
 \end{figure}

\subsection{Power law decay with  different strengths}

Following the same prescription as in the previous one, namely, exponential decay of interaction strengths, we now analyse the effect of external magnetic field on time-averaged entanglement between the subsystems, when the spin-spin interaction follows the power-law decay. Along with $\mathcal{Z}$, $\Delta$, and different pairs of spin in this scenario, we also have different strengths of interaction due to $\alpha_{p}$. We start presenting our results with varying \(\alpha_p\) which was absent in the previous case.

\begin{enumerate}




\item \emph {Effect of $\alpha_{p}$ on short and long-range entanglement. } Let us first consider the entanglement after averaging over time of the evolution for nearest neighbor (in Fig . \ref{fig:45vsa}). Like exponential decay case, if the initial state is prepared with low values of $\lambda =a$, time-averaged entanglement is higher than the case with high initial magnetic field for all values $\alpha_{p}$ as shown in Figs. \ref{fig:45vsa} and \ref{fig:47vsa}. Moreover, we  observe that for a fixed amount $\Delta$ and \(a\), if we keep on increasing the value of $\alpha_{p}$, the numerical value of \(\mathcal{L}_{avg}\) of \((4, 5)\)-pair is also increasing compared to lower value of $\alpha_{p}$. Most strikingly,
it is readily observed that higher value of $\alpha_{p}$ diminishes the effect of variable interactions on NN entanglement, i.e., the effect of $\mathcal{Z}$ is essentially wiped out with the increase of \(\alpha_p\), and we  get different averaged entanglement for different values of \(\Delta\) with the variation of \(a\) for any values of \(\mathcal{Z}\) (see right panel of  Fig. \ref{fig:45vsa}). 

 \begin{figure}[h]
\includegraphics[width = 8.6 cm]{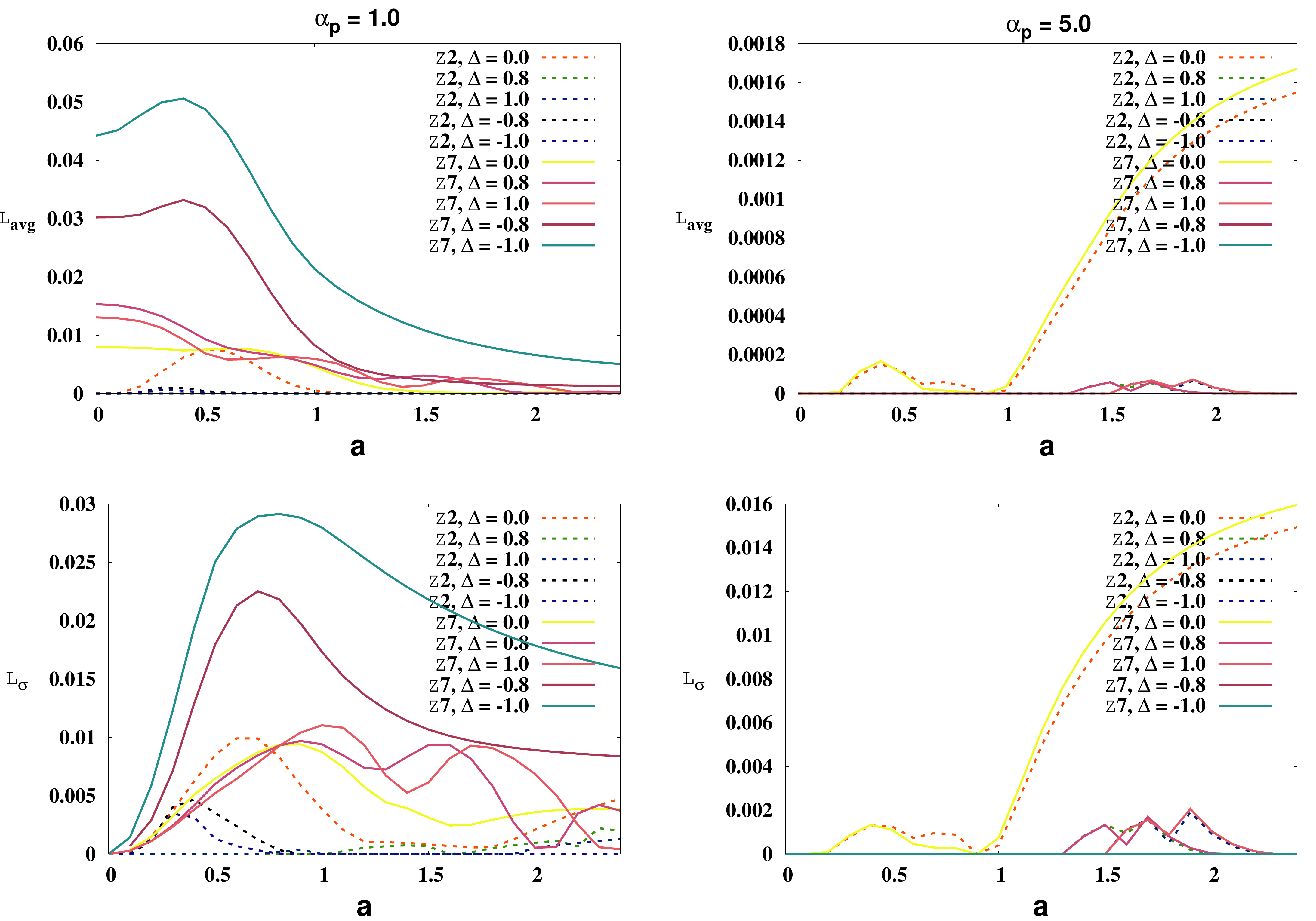}
\caption{(Top panel) Variation of time-averaged entanglement for \(\rho_{47}\) and (bottom panel) the respective fluctuations in entanglement with \(a\).  All other configurations are same as in Figs. \ref{fig:45vsa}. }
\label{fig:47vsa}
\end{figure}

On the other hand, long-range time-averaged entanglement shows a drastic change  with \(\alpha_p\) both in the qualitative and quantitative sense, as depicted in Fig. \ref{fig:47vsa}. First,    the numerical value of \(\mathcal{L}_{avg}\)  decreases a significant amount with the increase  of $\alpha_{p}$. Note that as one expects, long-range entanglement content is typically less than the NN ones.   
If we concentrate on Fig . \ref{fig:47vsa}, it is an immediate observation that the higher value of $\alpha_{p}$ is actually helping to generate higher value of long-range  time-averaged entanglement for high $a$, which is a basic qualitative difference from nearest neighbor scenario.
However, the diminishing effect of $\mathcal{Z}$ due to the increment of $\alpha_{p}$ is still present in the long-range scenario.  

%
%

\item \emph {Role of $zz$ interaction strengths. } As we observed, the effects of variable-range interactions wash out with the increase of \(\alpha_p\). For high \(\alpha_p\), we clearly see that \(\mathcal{L}_{avg} (\rho_{45})\) increases with the increase of \(\Delta\), both in the ferromagnetic and antiferromagnetic regimes, especially when the initial magnetic field is of moderate strength, i.e., when \(a <1\).  On the other hand, for small \(\alpha_p\),  the impact of  $\Delta$ for various values of $\mathcal{Z}$ is prominent and he behavior of time-averaged entanglement is qualitatively similar to the one described in the exponential case. For any \(\mathcal{Z}\), we find that moderate amount of ferromagnetic as well as antiferromagnetic interactions in the \(z\)-directions can generate high amount of NN time-averaged entanglement for \(a <1\) although the fluctuations of entanglement with time also increases with \(a\) and increase of \(\Delta\). However,  high content of average entanglement in the time-evolved state, \(\rho_{47}\) is obtained only when \(\Delta\) vanishes when the interactions is restricted to NN as well as NNN while presence of \(Delta\) gives benefit when we starts evolution with the system of long-range XYZ model with a moderate amount of magnetic field.

 \begin{figure}[h]
 \includegraphics[width = 8.6 cm]{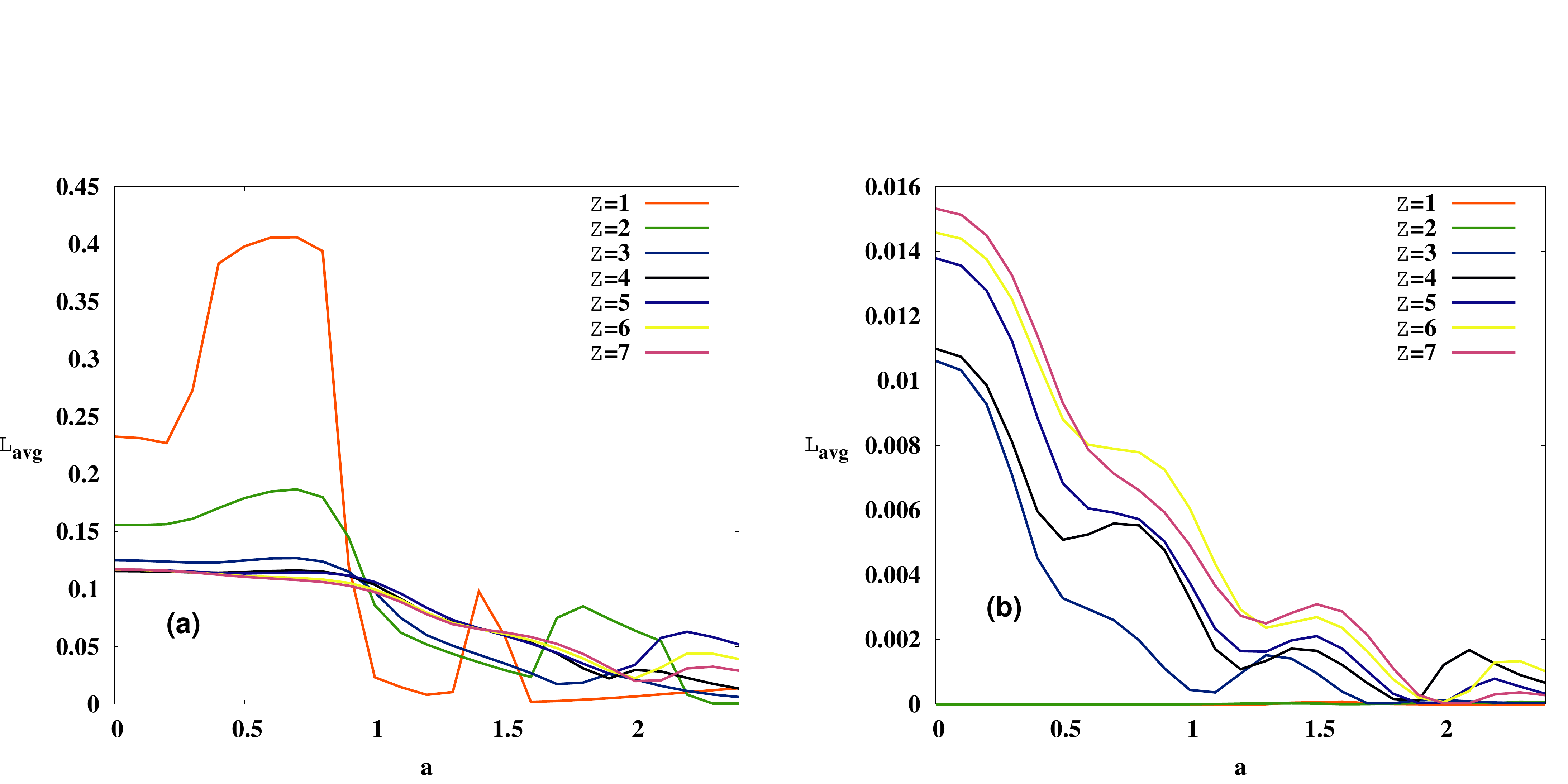}
\caption{ $\mathcal{L}_{avg} (\rho_{45})$ vs. \(a\) for different values of  $\mathcal{Z}$.  Here  $\Delta = 0.8$.  All other parameters and conditions are same as in Figs. \ref{fig:45vsa}.  }
\label{fig:entvsZ}
\end{figure}

\item \emph {Consequences of variable-range interactions on entanglement-dynamics. } As argued before, the beneficiary role of variable-range of interactions is only visible when the strength of  the decay is small. Universal feature observed in this scenario is that in presence of high magnetic field, entanglement content in the nearest neighbor pair  with different \(\mathcal{Z}\) values is almost  constant and is very low while when the initial strength of the magnetic field is  small to moderate,  \(\mathcal{L}_{avg} (\rho_{45})\) is maximum for the Hamiltonian having nearest neighbor interactions. The picture is completely opposite when we consider long-range entanglement, i.e., entanglement in \(\rho_{47} (t)\). Specifically, no entanglement is generated with the XYZ model having NN interactions and NN as well as NNN interactions. On the other hand, with long-range interactions of power-law decay can produce a high amount entanglement on average in the \((4,7)\)-pair and the amount of entanglement  in this pair decreases  with \(\mathcal{Z}\) and with the increase of \(a\), almost monotonically as shown in Fig. \ref{fig:entvsZ}.


\end{enumerate}

 \begin{figure}[ht]
\includegraphics[width = 8.0cm, height = 4.0cm]{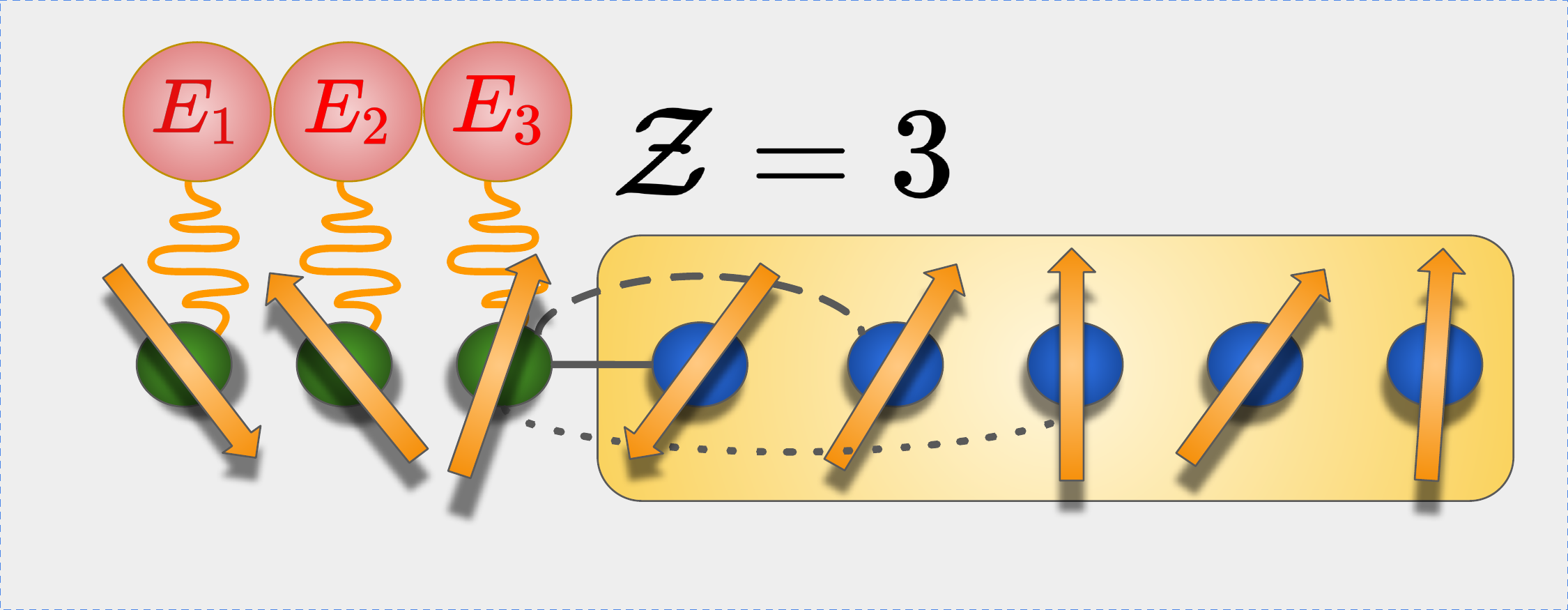} 
\caption{Schematic depiction of protecting spins from environment. Suppose there are eight interacting spins, following the Hamiltonian having variable-range of interactions, given in Eq. (\ref{eq:ham1}) with \(\mathcal{Z}=3\). Among them, three spins from the left are affected by the external noisy environments, denoted by \(E_i,\, i=1, 2, 3\) while the rest of the spins  is not. We show that in this  scenario, entanglement between pairs having different range can keep their initial value for a certain period of time. Therefore, the set-up of five spins which are not attached to the bath  forms a decoherence-free subspace by  prohibiting the effect of decoherence for a certain period of time. }
\label{fig:schematic}
\end{figure} 
 
\section{Preservation of entanglement in presence of noisy environment: Decoherence-free subsystems} 
\label{sec:open}

 \begin{figure*}[ht]
 \includegraphics[width = 18cm]{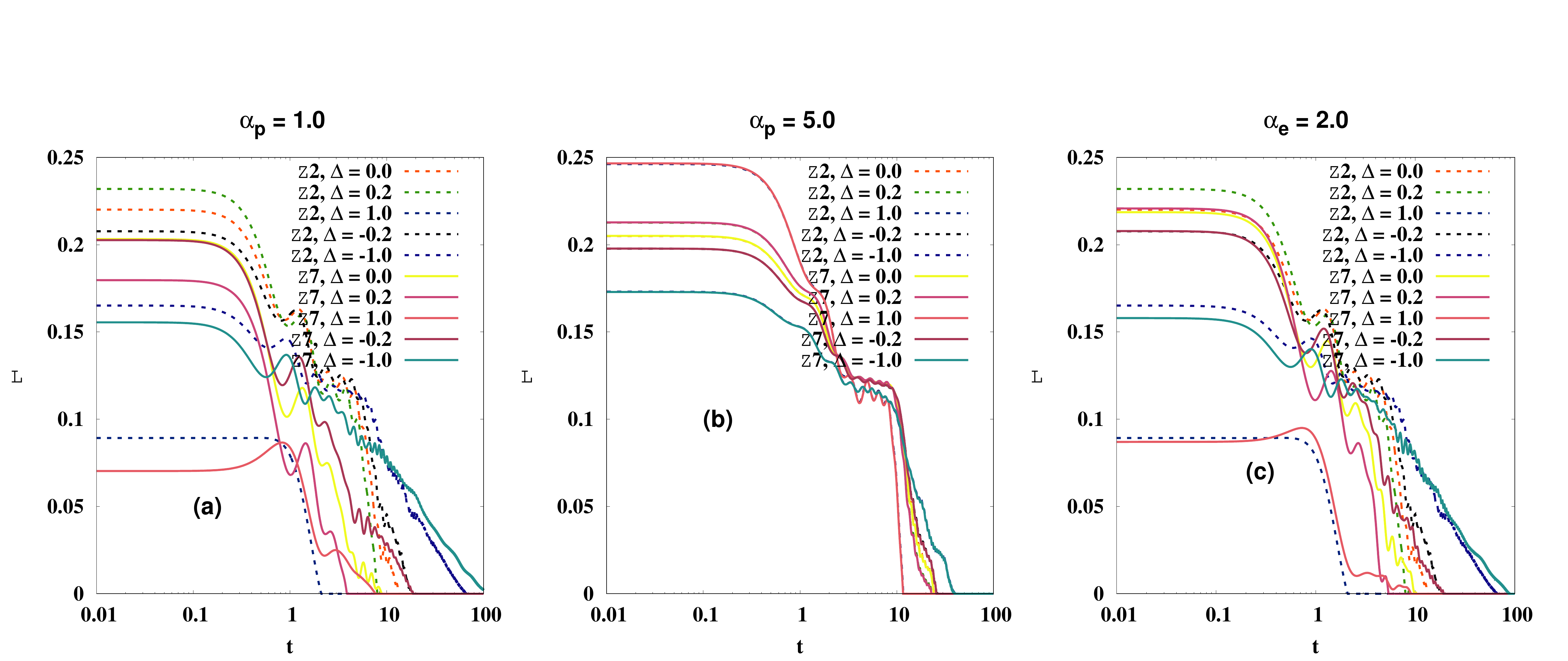}
 \caption{(From left to right)  $\mathcal{L} (\rho_{45})$ (ordinate) against $t$ (abscissa). The initial state is prepared in  the thermal state of the XYZ model with varying interaction strengths, \(\mathcal{Z}\) in presence of strong uniform magnetic field, \(\lambda=2.4\). \(\beta =20\).  Different lines correspond to different values of \(\Delta\) with power-law decay \(\alpha_p=1\), \(\alpha_p= 5\) and the exponential decay having \(\alpha_e=2\).  Dashed lines are for \(\mathcal{Z}=2\) and solid lines are for \(\mathcal{Z}=7\). The system consists of eight spins and the first three spins interact with bath according to repetitive interactions.  Here \(N=8\), and \(\gamma =0.8\). Although we have seen the change of \(\beta\) upto a very small value and \(\gamma\) do not qualitatively alter the results.  Entanglement is in ebits while time is dimensionless.   }
 \label{fig:45entvst}
 \end{figure*}

Preparing an isolated system is an ideal situation. Typically, the system prepared in a certain state starts interacting with environment, the quantum properties slowly decay, and the precious resource of entanglement vanishes with time. The question that is pertinent at this point ---is there a way to preserve quantum properties of the system or part of the system even in presence of system-environment interaction? Over the last few years, such questions are addressed and for  certain scenarios, protection mechanisms of bipartite, especially nearest neighbor entanglement are  developed. Here we will  propose a method which can keep entanglement as it is for a certain period of time (see schematic Fig. \ref{fig:schematic} for the depiction). 
 
 
 Let us consider a situation, involving  $N$ interacting spins and among them,  only $k$ spins are affected  by the environmental noise. For a nearest-neighbor Hamiltonian, it was shown \cite{titfreeze} that the entanglement of the remaining $N-k$ spins  can remain constant over a  certain amount of time since via LR bound, the noise to reach all the spins takes some time. In case of spin model with variable-range interactions, such a scenario is more involved and we can show that not only short-range entanglement, long-range entanglement can also freeze with a certain period of time.  For illustrative purposes, three spins of a total eight spins are chosen to be in contact with baths in a Markovian regime \cite{noise1, noisebook} and the rest five spins can be shown to act as a decoherence-free subspace, where decay of entanglement is slow. Analysis below considers the first three spins are attached to the bath and therefore, the patterns of entanglement in \(\rho_{(4,4+r)}, \, (r =1,2,3,4)\) are under study. Notice that if we consider \(\rho_{4,4-r},\,, r=1,2\), the entanglement shows collapse and revival for a certain periods of time.  
 
\begin{figure}[ht]
 \includegraphics[width = 8cm]{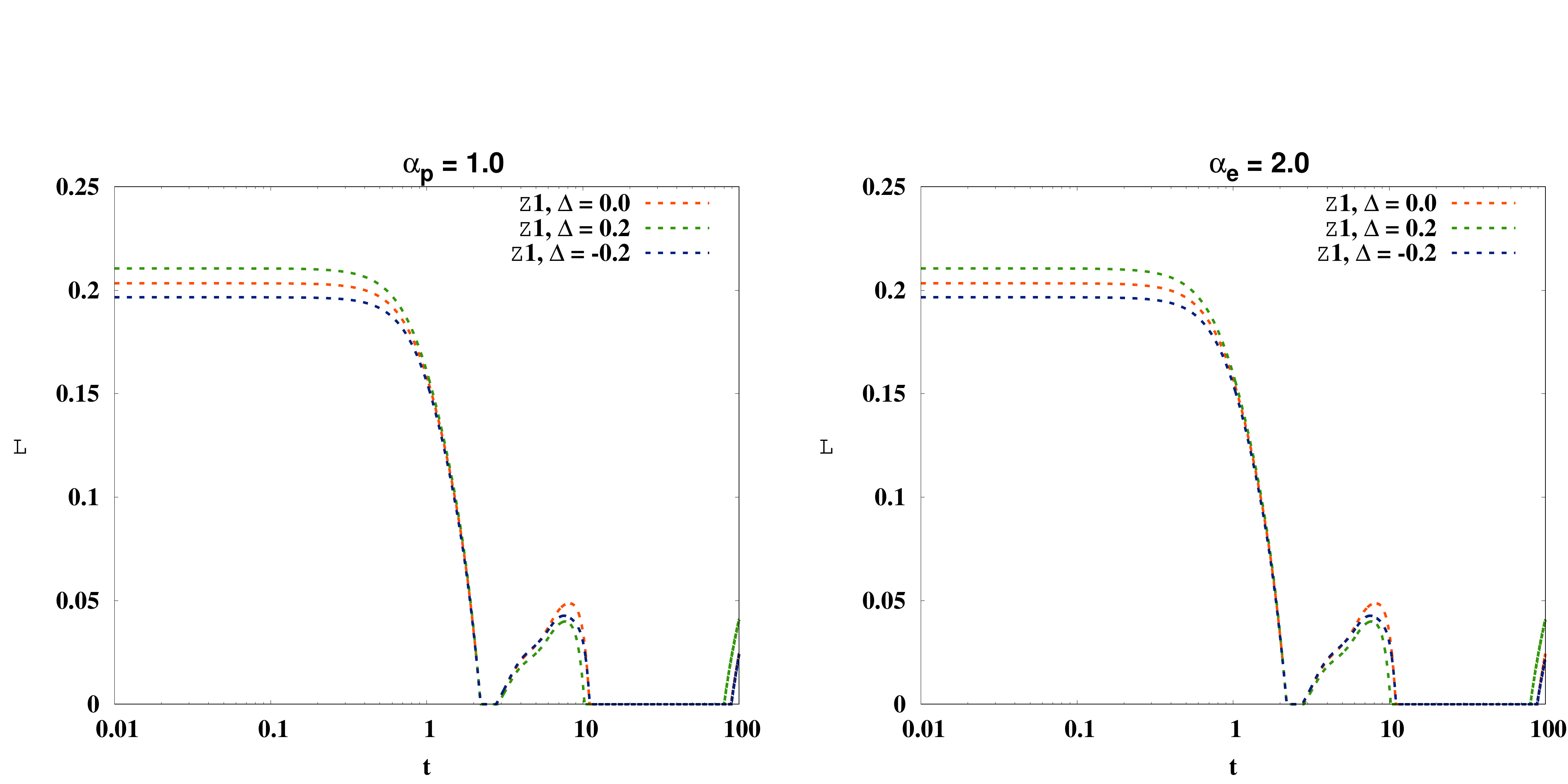}
 \caption{NN entanglement (vertical axis) vs. time (horizontal axis) for bosonic channel along with local dephasing in the \(z\)-direction. All the configurations are same as in Fig. \ref{fig:45entvst}. Left plot is for powerlaw decay with \(\alpha_p =1\) while the right one is for the exponential-decay case. Here \(s =0.5\). Unlike repetitive interactions,  entanglement is showing revival with time after the collapse.  Entanglement is in ebits while time is dimensionless.   }
 \label{fig:45entvstbosonic}
 \end{figure} 
 
%


 \subsubsection*{Freezing of entanglement}
 
 

Before moving further,  let us define the \emph{freezing of Entanglement} \cite{titfreeze}. Suppose LN of a pair of spins $(i, j)$ at $t$ is given as $\mathcal{L}(\rho_{(i,j)}(t))$, and its initial value at the beginning of the evolution can be referred as  $\mathcal{L} (\rho_{(i,j)} (t=0))$. An entanglement between \((i,j)\)-pair  is said to be frozen if 
 \begin{equation}
  | \mathcal{L}(\rho_{(i,j)}(t=0)) - \mathcal{L}(\rho_{(i,j)}(t))| < \delta,
 \end{equation}
 i.e., the derivative of the difference with respect to time vanishes.
 The duration of time, after which the entanglement starts changing with time is denoted by $\tau_f$ and can be called 
 the \emph{freezing terminal}. In other words, \(t >\tau_f\), the derivative in the difference between entanglement values at two neighboring instance of time is nonvanishing. 
 Precisely, entanglement values do not show any decay until $\tau_f$ upto the numerical accuracy of $\delta$. In our calculations,  we take $\delta = 10^{-5}$. The entanglement content of \(\rho_{(i, j)}\) when the freezing occurs, i.e., the entanglement value at \(\tau_f\) can be called frozen entanglement and is denoted by \(\mathcal{L}_f\).

\begin{figure*}[ht]
 \includegraphics[width = 18 cm]{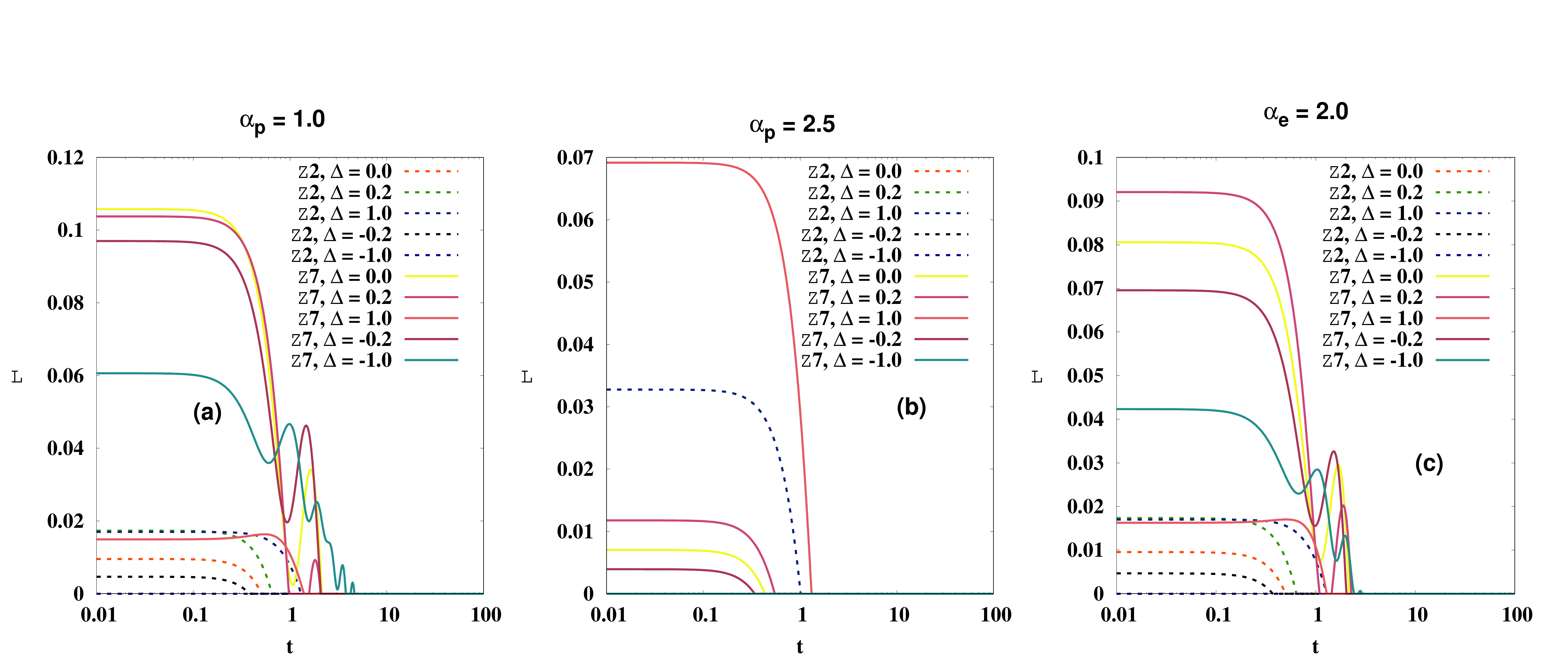}
 \caption{ $\mathcal{L} (\rho_{47})$ with $t$. All the configurations are same as in Fig.  \ref{fig:45entvst}. }
 \label{fig:47entvst}
 \end{figure*}

Before moving to the freezing terminal and frozen entanglement values, let us first discuss how the dynamics of entanglement is effected by the system parameters, \( \Delta, \lambda\) for a fixed \(\gamma\), the long-range interaction measured by \(\mathcal{Z}\), and \(\alpha_p\) for the power-law decay. In all these situations, the general trade-off that entanglement of any pairs show is that entanglement initially freezes for a certain period of time and then decay, as depicted in  Figs. \ref{fig:45entvst} and \ref{fig:47entvst} . However, unlike Hamiltonian with nearest neighbor case \cite{titfreeze},  the decay is not smooth, having lots of jiggling with time. Surprisingly, for a specific values of \(\Delta, \lambda\)-pair, we observe that $\mathcal{L}_{4,4+r}$  \((r=1,2 3, 4)\) can show a non-monotonicity with respect to time. In particular,  there is a certain time period at which the value $\mathcal{L}$ at $t > 0$ is more than that of \(t=0\) or the frozen value of entanglement. It is counter-intuitive in the sense, that it is usually assumed that  entanglement should reduce with time under the influence of environment. The overall trends of entanglement remains same for both repetitive interaction and bosonic baths. Only stark difference that we observe  is that when only bosonic absorption bath is active, after the initial freezing, entanglement pair collapses but it again revives with a certain period of time (see Fig. \ref{fig:45entvstbosonic}) for comparison between  entanglement dynamics with repetitive and bosonic environments). Let us now present the dependence of the parameters on all these observations.

 \emph{Dependence of dynamics on the strength of the decay in interactions. } As we have reported  in case of closed dynamics, high fall-off rate  removes the the \(\mathcal{Z}\) dependence from $\mathcal{L}$ even in the open systems. However, after freezing, the fidget behavior of entanglement gets smoother with the increase of \(\alpha_p\), thereby exhibiting much sharp decay towards collapse of entanglement (comparing left and middle Figs. \ref{fig:45entvst} and \ref{fig:47entvst}). Moreover, when we consider long-range interacting Hamiltonian, we expect to generate long-range entanglement which turns out to be false for high \(\alpha_p\). To illustrate other characteristics in dynamics, we again fix \(\alpha_p\) for presenting further observations.
%

 \emph{Role of $(\Delta, \lambda, \mathcal{Z})$  in freezing. } For a fixed anisotropy, the interplay of \(\Delta, \lambda\)-pair dictate the dynamics of the entanglement between any two-qubits  irrespective of the bath considered in this paper. It is interesting to point out here that there exists a surface in which bipartite entanglement vanishes in this model and it divides the plane into two regions having different entanglement patterns \cite{lakkaraju2020distribution}. Keeping this in mind as well as the observation in unitary dynamics in the preceding section, we choose \(\lambda\)    to be weak as well as strong which can manifest the the difference between the nearest neighbor model and the one having 
variable-range interactions.  In particular, for a nearest neighbor model, in presence of strong magnetic field, high delta leads to high amount of nearest neighbor entanglement while for weak  \(\lambda\), \(|\Delta|\) matters to obtain a good content of nearest neighbor entanglement. Such a universal scenario cannot be captured in the variable-range Hamiltonian since the interplay between \(\mathcal{Z}\), \(\Delta\) and \(\lambda\) is more sophisticated. In particular, ferromagnetic and the antiferromagnetic interactions in the $z$-direction behave differently for the long-range model, in presence of weak or strong magnetic field. 
 For example, we observe that for \(\mathcal{Z} \ne 1\), with high \(\lambda\) values, say, \(\lambda =2.4\),  $\Delta = 0.2$ is giving higher value of frozen entanglement  while when \(\lambda =0.4\), \(\Delta = -1\) can produce maximum frozen entanglement in any arbitrary pair, \(\rho_{4r}\). Moreover, among all the situations, ferromagnetic  long-range interacting XYZ model having low magnetic field turns out to be the most beneficial in overcoming the decoherence effects in the sense of generating high entanglement. Such observations possibly suggest that in the creation of long-range entanglement, the ratio between \(|\Delta|\) and \(\lambda\) in the long-range model  plays a role, i.e., when \(\Delta\) is high, \(\lambda\) should be low and vice-versa. The trends of frozen entanglement content and freezing terminal with the increase of \(\mathcal{Z}\) becomes clearer in the respective discussions below.

 \begin{figure}[h]
\includegraphics[width = 8cm]{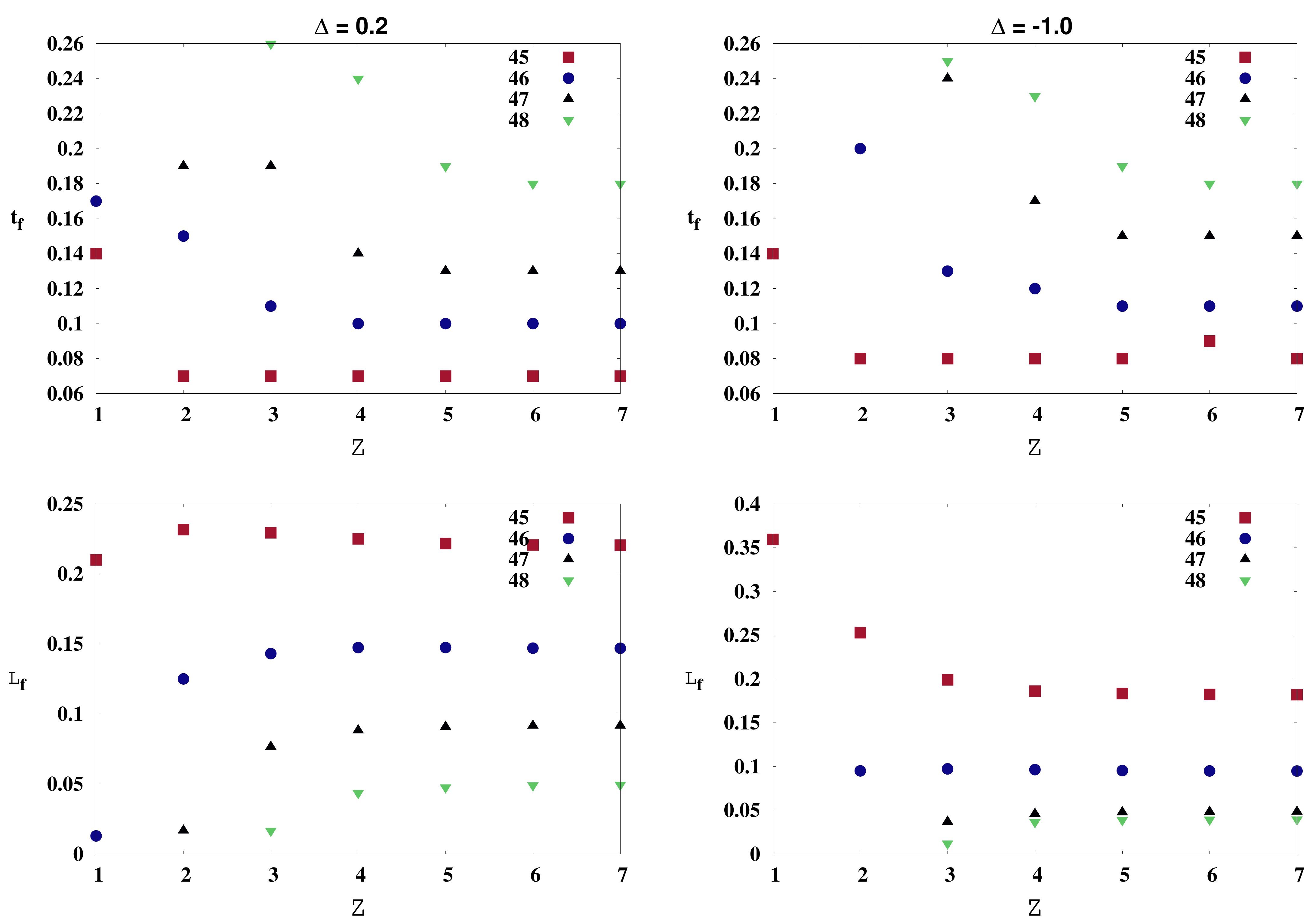}
\caption{
 (Upper panel) Freezing terminal  (vertical axis) vs. \(\mathcal{Z}\) (horizontal axis) while  frozen entanglement, \(\mathcal{L}_f\) (\(y\)-axis) with \(\mathcal{Z}\) (\(x\)-axis) (in Lower panel). Left column is for \((\Delta, \lambda) =(0.2, 2.4)\) while the right one is for the pair \((\Delta =-1, \lambda =0.4)\).  
The system is same as in Fig.  \ref{fig:45entvst} when the variable-range interactions decay exponentially. Different symbols correspond to different pairs of density matrices, \((4, 4+r)\).
 It is clear that \(\tau_f\) decreases with \(\mathcal{Z}\) while the opposite picture emerges for frozen entanglement content. The maximum value of \(\tau_f\)  is obtained for a minimum \(\mathcal{Z}\) required to make the entanglement of the corresponding  density matrices nonvanishing. On the other hand, except nearest neighbor frozen entanglement value, \(\mathcal{L}_f\) increases and saturates with \(\mathcal{Z}\). All the axes are dimensionless in the upper panel while the vertical axis in the lower panel is in ebits.
 }
\label{fig:tfvsZLfvsZexp}
\end{figure}

 \begin{figure}[h]
\includegraphics[width = 8cm]{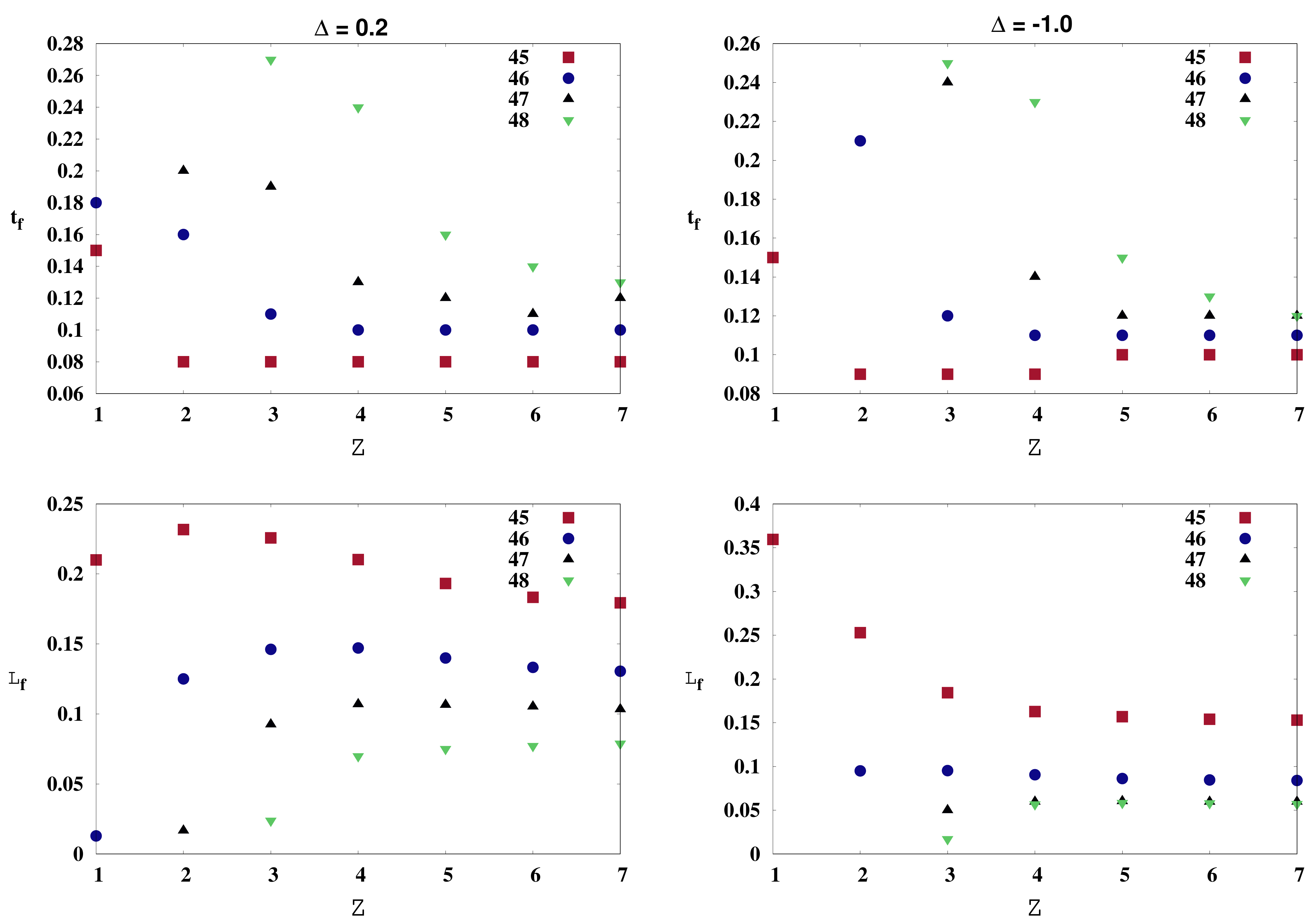}
\caption{Freezing terminal and frozen entanglement with \(\mathcal{Z}\) for Coulomb law. All the configurations are  same as in Fig. \ref{fig:tfvsZLfvsZexp}.  }
\label{fig:tfvsZLfvsZpower}
\end{figure}

 \subsection{Impact of variable-range  interactions on freezing terminal: Complementarity between terminal and frozen value } 
 
 Independent of baths, in the decoherence-free scenario, entanglement in \(\rho_{4,4+r}\) always show freezing provided the initial Hamiltonian is long-range. First observation in this case is that for a fixed \(\mathcal{Z}\), \(\tau_f\) obtained for  \(\rho_{45}\) is smaller than  \(\rho_{46}\) and so on, i.e., we find that 
 \begin{eqnarray}
 \tau_f (\rho_{45}) < \tau_f (\rho_{46}) < \tau_f (\rho_{47}) < \ldots. 
\label{eq:tauf}
\end{eqnarray}  
It can possibly be argued that  the effects of decoherence can reach the sites closer to the bath faster than that of the further. As we will show now, such a simple argumentation of information flow cannot explain all the results.   

Secondly, we find that \(\tau_f\) for \(\rho_{45}\) does not change at all with the increase of \(\mathcal{Z}\) although for long-range case, it slowly decreases with \(\mathcal{Z}\) and saturates to a certain value. It demonstrates that the value of \(\mathcal{Z}\) which is enough to generate the long-range entanglement leads to the maximum freezing terminal. After that value,  increasing \(\mathcal{Z}\) has an adverse  effect on freezing terminal. It implies that increasing interactions possibly induces more system-bath interaction, thereby making the decoherence effects prominent.   Specifically, since \(\mathcal{Z}=2\) is sufficient to generate entanglement in \(\rho_{46}\) while we require higher than NNN interactions, i.e., \(\mathcal{Z}>3\),  to have nonvanishing entanglement in \(\rho_{47}\) and \(\rho_{48}\), the maximum freezing terminal is obtained for \(\mathcal{Z}= 2\) in case of  \(\rho_{46}\) and \(\mathcal{Z}=3\) in case of \(\rho_{47}\) and \(\rho_{48}\), thereby pinpointing a critical value of \(\mathcal{Z}\) leading to highest freezing terminal. 
For illustration and motivated from the previous findings, we choose \(\Delta, \lambda\)-pair as \((-1, 0.4)\) and \((0.2, 2.4)\) as shown in the upper panels of  Figs. \ref{fig:tfvsZLfvsZexp} and \ref{fig:tfvsZLfvsZpower}.  Such a detrimental effect of \(\mathcal{Z}\) in a freezing terminal can be compensated from the frozen entanglement value as will be shown now.

 \textit{Complementary relation between $\mathcal{L}_f$ and $\tau_f$.} We find that the frozen entanglement value and freezing terminal obeys a complementary value for a fixed \((i,j)\)-pair. We propose that for a given \(\rho_{ij}\), 
 \begin{eqnarray}
  \mathcal{L}_f  + \tau_f \leq c,
 \end{eqnarray}
 where \(c\) depends on the values of \(i\) and \(j\). From our analysis, we find that it never goes beyond \(0.35\) which is obtained for the nearest neighbor case. Note that under unitary dynamics, the model can at most generate nearest neighbor entanglement around this value. At this point, it is reasonable to conjecture that \(c\) cannot go beyond the average entanglement that can be created in the system under different quenches without decoherence, i.e.,  \( c \leq \mathcal{L}_{avg} (\rho_{(i,j)})\). 
 
 The existence of such a relation guarantees that the frozen entanglement content, \(\mathcal{L}_f\), increases with \(\mathcal{Z}\) although nearest neighbor entanglement decreases with \(\mathcal{Z}\) as depicted in the lower panels of  Figs. \ref{fig:tfvsZLfvsZexp} and \ref{fig:tfvsZLfvsZpower}. To hold complementarity, we find an opposite hierarchy among frozen value of entanglement than the one obtained for \(\tau_f\), in Eq. (\ref{eq:tauf}), i.e. for a fixed \(\mathcal{Z}\),  \( \ldots < \mathcal{L}_f (\rho_{48}) < \mathcal{L}_f (\rho_{47}) < \mathcal{L}_f (\rho_{46}) < \mathcal{L}_f (\rho_{45}) \).
 The dependence of the complementary relation for long-range entanglement on \(\mathcal{Z}\)  turns out to be highly nontrivial. Specifically, it has an optimal \(\mathcal{Z}\) value when it reaches the maxima, thereby reflecting the dependence  of \(\tau_f\) on \(\mathcal{Z}\) as discussed before.

 
%
 

\section{Conclusion}
\label{sec:conclusion}

Quantum spin models with variable-range interactions can exhibit certain characteristics like continuous symmetry breaking phase which cannot,  in principle, be seen by the corresponding short-range models and hence it is quite plausible that quantum technologies can be designed by exploiting them. On the other hand, entanglement generated via dynamical systems of the quantum spin models is beneficial for several quantum information protocols. However,  the systems with variable-range interactions  are,  in general, intractable analytically  which makes  the study of these models from the perspective of quantum information to be  limited.

In this work, we studied the patterns of entanglement produced in the evolved state of the anisotropic quantum XYZ model with a uniform magnetic field by varying interactions according to the exponential as well as the power-law decays. The quenching for dynamics is performed by tuning the magnetic field.   We found that the maximum amount of short-range as well long-range time-averaged entanglement can be created in the presence of both ferromagnetic as well as antiferromagnetic couplings in the $z$-direction with a moderate amount of power-law and exponential interactions. Moreover, we showed that the high content of bipartite entanglement in dynamics can be  established at the cost of high fluctuations. 

When the part of the thermal state of the same model came to  contact with environments which are modeled by the repetitive interactions and bosonic bath along with dephasing channels, the bipartite entanglement of the rest of the part remains constant at the beginning of the evolution before decaying to vanish. We identified the regions in the parameter space which can be tuned to obtain the maximum time when entanglement remains constant. We also reported that frozen entanglement value and the freezing time follow a complementary relation. Although all the observations  remain qualitatively the same for both the baths, we found that for the bosonic baths in presence of dephasing noise in the $z$-direction, along with the freezing of entanglement, entanglement also shows a revival after the collapse which is not seen in the repetitive case. The investigations carried out in this paper indicate that the XYZ model with varying interaction strength which can be realized by using currently available technology is the potential candidate for building  quantum devices even in presence of  the noisy environment. 
 
\section*{Acknowledgement}
We acknowledge the support from Interdisciplinary Cyber Physical Systems (ICPS) program of the Department of Science and Technology (DST), India, Grant No.: DST/ICPS/QuST/Theme- 1/2019/23. We  acknowledge the use of \href{https://github.com/titaschanda/QIClib}{QIClib} -- a modern C++ library for general purpose quantum information processing and quantum computing (\url{https://titaschanda.github.io/QIClib}) and cluster computing facility at Harish-Chandra Research Institute.

\section*{Appendix} 
 
%
  
In this paper, the entire analysis has been carried out under this Markovian approximation where the  evolution of the system is governed by the Gorini-Kossakowski-Sudarshan-Lindblad (GKSL) master equation \cite{noise1, noisebook}, given by
   \begin{equation}
\frac{d \rho_{S}}{d t}=-\frac{i}{\hbar}\left[H_{S}, \rho_{S}\right]+\mathcal{D}\left(\rho_{S}\right),
\label{lindblad}
\end{equation}
where $H_S$ represents the system's Hamiltonian and $\mathcal{D}$ is the dissipative part dictated by the choice of the environment. In our case, it is fixed by  the local repetitive baths and bosnoic environment. 

\subsection{Local repetitive baths}

Consider a bath consists of collection of baths,  $E_i$, \((i=1, 2, \ldots, n)\) which are not interacting with each other and whose Hamiltonian is given by $H_{E_{i}}=B \sigma_{i}^{z}$. Each of those spins are interacting locally with spins $\sigma_i$ of the system for a certain short-period of time. The total evolution time can be divided into small time-intervals, \(0, \delta t, 2 \delta t \ldots\). The system interacts with the environment during \((0, \delta t)\),  with the interacting Hamiltonian, given by \cite{dhahri2008lindblad, PhysRevLett.102.207207, PhysRevA.91.040303, attal2006repeated, PhysRevA.34.1642, barchielli1991measurements}
\begin{equation}
H_{\mathrm{int}}(\delta t)= \sum_i \sqrt{k / \delta t}\left(\sigma_{i}^{x} \otimes \sigma_{E}^{x}+\sigma_{i}^{y} \otimes \sigma_{E}^{y}\right),
\end{equation}
where $k$ has the dimension of $\text(energy)^2 \times \text(time)$ and $i$ is the index of spin in the system. After \(\delta t\), the system-environment pair becomes entangled and we reset the environment to the initial thermal state and system's part is obtained by tracing out environment form the entangled state.   The final Hamiltonian of the system-environment i can be represented as 
\begin{equation}
H=H_{S} \otimes \mathbb{I}_{E}+\mathbb{I}_{S} \otimes H_{E}+H_{\mathrm{int}}(\delta t)
\end{equation}
In our work, the three spins of the system  are interacting with the bath, which leads to the $\mathcal{D}(\rho_S)$ of the form 
 \begin{equation}
\mathcal{D}\left(\rho_{S}\right)=\frac{2 k}{\hbar^{2}}\sum_{i=1}^{3} \sum_{l=0}^{1} p_{l}\left[2 \eta_{i}^{l+1} \rho_{S} \eta_{i}^{l}-\left\{\eta_{i}^{l} \eta_{i}^{l+1}, \rho_{S}\right\}\right]
\label{repetative}
\end{equation}
where	$p_{l}=Z_{E}^{-1} \exp \left[(-1)^{l} \beta_{E} B\right]$, $Z_{E}=\operatorname{tr}\left[\exp \left(-\beta_{E} H_{E}\right)\right]$ and $\eta_{d_{i}}^{\alpha}=(\sigma_{d_{i}}^{x}+i(-1)^{\alpha} \sigma_{d_{i}}^{y})/2$. 

The initial state is considered to be the thermal state  of the XYZ model, given in Eq. (\ref{eq:ham1}) with $\beta_S = 20$ and then following the evolution in Eq. (\ref{lindblad}), at each time interval, we trace out all the parties except \(4\) and \(k\) \((k =5, \ldots, 8)\), and calculate LN of $\rho_{(4,k)}$.  
 Since Eq.  (\ref{lindblad}) is a differential equation, we use fourth order Runge-Kutta method to solve the equation, whose numerical errors accumulate in the fifth power of time increment. We use $\delta = 0.01$ so that error is of the order $10^{-5}$.\\

\subsection{Bosonic environment}

Let us consider another decoherence model where the system is permanently connected to the local thermal baths of harmonic oscillators for a particular interval of time \cite{noise1, PhysRevLettbosonic, CHANDA20161}. The particular  sub-systems are also connected to local bosonic reservoirs  which act as  absorption channels, thereby helping to accumulate energy from the environment to the system. We also apply local bosonic reserioirs after a certain interval of time to extract energy from  it, that means, those acts as a dissipation channel. In all through the process, a local dephasing noise which is either in the $x$ direction or in the $z$ direction is acting on the selected parties of the system.
Suppose a single qubit is exposed to the dephasing noise, then the total Hamiltonian reads
\begin{equation}
H = \omega_{0}\sigma_{z} + \sum_{i} a_{i}^{\dagger}a_{i} + \sum_{i} \sigma_{noise}(g_{i}a_{i} + g_{i}^{*}a_{i}^{\dagger}),
\end{equation}  
where $a_{i}$ and $a_{i}^{{\dagger}}$ are the annihilation and creation operators for the mode  $i$ of the harmonic oscillators. $\omega_{0}$ is the energy spacing in the qubit, $g_{i}$ is the coupling constant between reservoir and qubit while  $\omega_{i}$ is the frequency for each mode. Here $\sigma_{noise}$ is either $\sigma_{x}$ or $\sigma_{z}$ depending on the direction of applied dephasing channel. By putting all these into the GKSL master equation, the resulting form of the noise in the dynamical part looks like
\begin{equation}
\mathcal{D}(\rho_{N}^{\beta}) = \gamma_{deph}(t)(\sigma_{noise}\rho_{N}^{\beta}\sigma_{noise} - \rho_{N}^{\beta}),
\end{equation}
where $\gamma(t)$ is the time-dependent dephasing rate which can be calculated from the spectral density of the reservior. As we consider the initial state of the system as the thermal state (with large value of $\beta$), and the reservior is characterized by the Ohmic spectral density $\mathcal{K}(\omega)$ \cite{haikka'13}, given by 
\begin{equation}
\mathcal{K}(\omega) = \frac{\omega ^{s}}{\omega^{s-1}_c} \exp(-\frac{\omega}{\omega_{c}}),
\end{equation}  
with \(\omega\)  and \(\omega_c\) being  respectively the frequency and the cut-off frequency  of the reservoir and \(s\) is the Ohmicity parameter,  the time-dependent dephasing rate, 
\begin{equation}
\gamma_{deph}(t,s) = (1+ (\omega_{c}t)^{2})^ {- \frac{s}{2}} \Gamma(s) \sin[s \tan^{-1}(\omega_{c} t)],
\end{equation}
where $\Gamma(s)$ is the Euler-gamma function. $s$ is the ohimicity parameter. Since the environments are independently acting on the chosen parties, the effect of local environments acting on, say, three spins of the systems can be achieved  by summing up three of them. Hence, the dynamical part  in this situation reduces to
\begin{equation} 
\mathcal{D}_d(\rho_{N}^{\beta}) = \sum_{i=1}^{d}\mathcal{D}_{d_{i}}(\rho_{N}^{\beta}).
\end{equation}

\bibliography{freezing.bib}

\end{document}